\documentclass{aa}  

\usepackage{graphicx}
\usepackage{txfonts}
\usepackage[colorlinks=true,citecolor=cyan]{hyperref}
\usepackage{subcaption}         
\usepackage{lscape}             
\usepackage{placeins}           
                                
\begin{document}

   \title{The CepA disk-outflow system at $\leq 0.2''$ or $\leq 100$\,au resolution}

   \subtitle{Northern Extended Millimeter Array (NOEMA) long baseline data}


   \author{H.~Beuther\inst{1}
        \and 
        C.~Gieser\inst{1}
        \and 
        V.~Aberham\inst{1}
        \and 
        J.M.~Winters\inst{2}
        \and 
        R.~Neri\inst{2}
        \and 
        A.~Ahmadi\inst{3}
        \and 
        R.~Kuiper\inst{4}
        \and 
        Th.~Henning\inst{1}
        \and 
        H.~Linz\inst{1}
        \and 
        T.~M\"oller\inst{5}
        \and 
        V.~Elbakyan\inst{4}
        \and 
        L.~Moscadelli\inst{6}
        \and 
        D.~Semenov\inst{1,7}
        \and 
        J.~Urquhart\inst{8}
        \and 
        P.~Klaassen\inst{9}
        \and 
        M.~Beltran\inst{10}
        \and 
        \'A.\ S\'anchez-Monge\inst{11,12}
        \and 
        T.~Peters\inst{13}
        \and 
        R.~Galvan-Madrid\inst{14}
        \and 
        S.~Leurini\inst{15}
        \and 
        S.~Lumsden\inst{16}
        \and 
        R.E.~Pudritz\inst{17}
        \and 
        A.~Palau\inst{14}
        \and 
        H.~Zinnecker\inst{18}
        }

   \institute{$^1$ Max Planck Institute for Astronomy, K\"onigstuhl 17,
     69117 Heidelberg, Germany, \email{beuther@mpia.de}\\
     $^2$ IRAM, 300 rue de la Piscine, Domaine Universitaire de Grenoble, 38406 St.-Martin-d’Hères, France\\
     $^3$ ASTRON, The Netherlands Institute for Radio Astronomy, Postbus 2, NL-7990 AA Dwingeloo, The Netherlands\\
     $^4$ Faculty of Physics, University of Duisburg-Essen, Lotharstraße 1, D-47057 Duisburg, Germany\\
     $^5$ I. Physikalisches Institut, Universität zu Köln, Zülpicher Straße 77, 50937 Cologne, Germany\\
     $^6$ INAF - Osservatorio Astrofisico di Arcetri, Largo E. Fermi 5, I-50125, Firenze, Italy \\
     $^7$ Institut f\"{u}r Theoretische Astrophysik, Albert-Ueberle-Str. 2, 69120 Heidelberg, Germany\\
     $^8$ Centre for Astrophysics and Planetary Science, University of Kent, Canterbury CT2 7NH, UK\\
     $^9$ United Kingdom Astronomy Technology Centre, Edinburgh, UK\\
     $^{10}$ INAF-Osservatorio Astrofisico di Arcetri, Largo E. Fermi 5, I-50125 Firenze, Italy\\
     $^{11}$ Institut de Ci\`encies de l'Espai (ICE), CSIC, Campus UAB, Carrer de Can Magrans s/n, E-08193, Bellaterra, Barcelona, Spain\\
     $^{12}$ Institut d'Estudis Espacials de Catalunya (IEEC), E-08860, Castelldefels, Barcelona, Spain\\
     $^{13}$ 85238 Petershausen, Germany\\
     $^{14}$ Universidad Nacional Aut\'onoma de M\'exico, Instituto de Radioastronom\'ia y Astrof\'isica, Antigua Carretera a P\'atzcuaro 8701, Ex-Hda. San Jos\'e de la Huerta,\\ 58089, Morelia, Michoac\'an, M\'exico\\
     $^{15}$ INAF - Osservatorio Astronomico di Cagliari, Via della Scienza 5, 09047 Selargius (CA), Italy\\
     $^{16}$ School of Physics and Astronomy, The University of Leeds, Wood- house Lane, Leeds LS2 9JT, UK\\
     $^{17}$ Department of Physics and Astronomy, McMaster University, 1280 Main St. W, Hamilton, ON L8S 4M1, Canada\\
     $^{18}$ Nucleo de Astroquimica y Astrofisica, Universidad Autónoma de Chile, Av. Pedro de Valdivia 425, Santiago, Chile
}

   \date{\today}

 
  \abstract
   {Although there has been significant progress, the physical properties and potential fragmentation of accretion disks around high-mass protostars remain poorly constrained.}
   {We characterize at high angular resolution one of the most nearby ($\sim$700\,pc) high-mass accretion disk candidates CepA HW2.} 
   {Using the new long baseline array configuration ($\sim$1700\,m) of the Northern Extended Millimeter Array (NOEMA), we study CepA HW2 with a resolution of $\leq$0.2$''$ or $\leq$100\,au at 1.3\,mm in dust continuum and spectral line emission.}
   {The mm continuum emission resolves the central disk candidate into several sub-structures. Conducting a Toomre $Q$ stability analysis based on CH$_3$CN and continuum data, and a comparison to 3D radiation hydrodynamic simulations shows that the data are consistent with an almost edge-on disk where the observed sub-structures may represent fragments within the disk. The CO and SiO spectral line data confirm a second bipolar outflow (in addition to the well-known jet) emanating from the central peak position. This indicates that this central peak should host at least a binary if not even a higher order multiple system. The usually assumed dense gas tracer CH$_3$CN shows also contributions from the outflows which complicates further kinematic analysis of the disk.}
   {The high-resolution outflow-disk data of CepA reveal a multiply fragmented disk that drives several outflows. These observations enforce the picture of high-mass star formation where multiplicity and fragmentation can happen on the smallest spatial scales related to the inner accretion disks.}


   \maketitle
\nolinenumbers

\section{Introduction}

Although there has been significant progress in understanding high-mass star formation (e.g., \citealt{zinnecker2007,beuther2006b,tan2014,motte2018,kumar2020,beuther2025}), we are still lacking good knowledge of the properties of the accretion disks around massive ($>$8\,M$_{\odot}$) protostars.  Larger-scale rotating structures, also termed toroids, have been reported regularly (e.g., \citealt{cesaroni2007,beltran2016}), however, real central high-mass accretion disks have still proven to be extremely rare and very hard to observe. A few candidates are Orion source $I$ \citep{ginsburg2018}, AFGL4176mm1 \citep{johnston2015,johnston2020}, G11.92-0.61 \citep{ilee2016,ilee2018}, IRAS16547-4247 \citep{zapata2019}, G17.64+0.16 \citep{maud2019}, G023.01-00.41 \citep{sanna2021} or W75N(B)-VLA2 and VLA3 \citep{gomez2023,sanchez-monge2025}. Recently, high-spatial resolution candidate sample studies have been presented in \citet{olguin2026} and \citet{yang2026}. Important open questions to be studied relate to their physical and chemical properties and how much the disks are prone to fragmentation to form multiple systems.

The target of this study, the disk candidate within Cepheus A, also called CepA HW2 \citep{hughes1984}, has been a high-mass disk candidate for many years (e.g., studies by \citealt{patel2005}, \citealt{comito2007}, \citealt{debuizer2017}, \citealt{beuther2018b}, \citealt{ahmadi2023} or \citealt{sanna2025}). At a distance of 0.7\,kpc \citep{moscadelli2009}, it is, after Orion source $I$, one of the closest high-mass star-forming regions and hence allows very high linear spatial resolution studies. Its bolometric luminosity is reported as $1.5\times 10^4$\,L$_{\odot}$ \citep{beuther2018b}. Early studies reveal a molecular outflow in the northeast-southwest direction \citep{gomez1999} that should be driven by a central jet traced in the ionized cm continuum emission \citep{torrelles1996,jimenez2007,carrasco-gonzalez2021}. The study by \citet{torrelles1996} also reveals H$_2$O maser emission, which was first proposed to stem from a disk perpendicular to the jet, but was later shown by VLBA proper motion observations to trace a wide-angle outflow \citep{torrelles2011}. CH$_3$OH class {\sc II} maser studies reveal infall and rotational motions \citep{sugiyama2014,sanna2017}, whereas polarized CH$_3$OH maser investigations indicate that the magnetic field regulates the accretion onto the disk \citep{vlemmings2010}. While \citet{patel2005} discuss a relatively large disk-like structure around a central 15\,M$_{\odot}$ object perpendicular to the mentioned outflow and jet, \citet{comito2007} report that this central rotating structure may contain at least three protostars, and that the actual accretion disk around the central peak may be on even smaller spatial scales. Cepheus A has also been one of the first regions where a molecular outflow had been identified, roughly in east-west direction \citep{rodriguez1980}. This structure is consistent with extended H$_2$ emission \citep{cunningham2009} and early small-scale SiO emission \citep{comito2007}. \citet{cunningham2009} also discuss that this east-west outflow maybe the earlier incarnation of the northeast-southwest outflow in the framework of a precessing jet system. \citet{martin-pintado2005} report the possibility of an intermediate-mass hot molecular core about $0.4''$ east of the radio jet and main continuum peak. Adding three suggested protostars by \citet{curiel2002}, a total of six potential protostars are discussed within the inner $1''$ of the region \citep{comito2007}.  The SOFIA massive star formation survey fitted the spectral energy distribution of the region, resulting in best fit solutions for the central object between 12 and 16\,M$_{\odot}$ \citep{debuizer2017}. CepA is also subject to a $\sim$5\,yr periodicity observed in CH$_3$OH maser emission and through near-infrared light echo analysis \citep{durjasz2022,stecklum2025}. A recent NH$_3$ study with the Very Large Array has found collapsing circumstellar gas with infall rates around $2\times 10^{-3}$\,M$_{\odot}$yr$^{-1}$ \citep{sanna2025}.

\begin{table}[htb]
\caption{Continuum and spectral line parameters.}
\begin{tabular}{lrrr}
  \hline \hline
line & freq. & $1\sigma$ & beam \\
& (GHz) & $\left(\frac{\rm mJy}{\rm beam}\right)$ & ($''$) \\
\hline
continuum & 225.498 & 0.36 & $0.17''\times 0.11''$ \\
SO$(7_8-7_7)$	                 & 214.357 & 3.76 & $0.20''\times 0.12''$ \\ NH$_2$D$(9_{6,4}-10_{3,8})$      & 214.991 & 3.89 & $0.20''\times 0.12''$\\   
KCl$(28-27)$	                 & 215.008 & 3.68 & $0.20''\times 0.12''$\\ 
SO$(5_5-4_4)$	                 & 215.221 & 3.60 & $0.20''\times 0.12''$\\ 
DCO$^+(3-2)$	                 & 216.113 & 3.53 & $0.20''\times 0.11''$\\    
D$_2$CO$(8_{1,7}-8_{1,8})$       & 216.492 & 3.75 & $0.20''\times 0.11''$\\   
NH$_2$D$(3_{2,2}-3_{1,2})$       & 216.563 & 3.67 & $0.20''\times 0.11''$\\    
SiO$(5-4)$                       & 217.105 & 3.78 & $0.20''\times 0.11''$\\ 
DCN$(3-2)$	                     & 217.239 & 3.47 & $0.20''\times 0.11''$\\    
HDCO$(8_{3,6}-9_{2,7})$          & 217.586 & 3.56 & $0.20''\times 0.11''$\\    
CH$_3$CN$(14_0-13_{-2})v8$	             & 217.606 & 3.59 & $0.20''\times 0.11''$\\   
HDCO$(4_{2,2}-5_{1,5})$          & 217.908 & 3.65 & $0.20''\times 0.11''$\\ 
H$_2$CO$(3_{0,3}-2_{0,2})$          & 218.222 & 3.33 & $0.20''\times 0.11''$ \\
HC$_3$N$(24-23)$                 & 218.325 & 3.40 & $0.20''\times 0.11''$\\ 
CH$_3$OH$(4_2-3_1)$              & 218.440 & 3.44 & $0.20''\times 0.11''$ \\
SO$_2(3_{2,2}-3_{1,2})v2=1$         & 219.466 & 3.35 & $0.20''\times 0.11''$ \\   
C$^{18}$O$(2-1)$	             & 219.560 & 3.19 & $0.20''\times 0.11''$ \\ 
SO$(6_5-5_4)$                    & 219.949 & 3.30 & $0.20''\times 0.11''$ \\
CH$_3$OH$(7_1-8_0)$              & 220.079 & 3.22 & $0.20''\times 0.11''$ \\
HCOOCH$_3(17_{4,13}-16_{4,12})$  & 220.167 & 3.26 & $0.20''\times 0.11''$ \\
CH$_2$CO$(11_{1,11}-10_{1,10})$  & 220.177 & 3.28 & $0.20''\times 0.11''$ \\  
$^{13}$CO(2--1)	                 & 220.399 & 3.21& $0.20''\times 0.11''$ \\ 
CH$_3$CN$(12_k-11_k)k=0$             & 220.747 & 3.44 & $0.20''\times 0.11''$ \\
CH$_3$CN$(12_k-11_k)k=...$           & ... \\
CH$_3$CN$(12_k-11_k)k=7$             & 220.539 & 3.44 & $0.20''\times 0.11''$\\
CH$_3$OH(3-2)	                 & 221.295 & 3.43 & $0.20''\times 0.11''$ \\    
CH$_3$CN$(12_2-11_2)v8$	             & 221.422 & 3.30 & $0.20''\times 0.11''$\\   
CH$_3$OH(3-4)	                 & 230.027 & 3.90 & $0.19''\times 0.11''$\\    
KCl$(30-29)$	                 & 230.321 & 3.67 & $0.19''\times 0.11''$\\     
CO$(2-1)$                        & 230.538 & 3.91 & $0.19''\times 0.11''$\\  
H30$\alpha^a$                    & 231.901 & 2.71 & $0.19''\times 0.11''$\\         
H$_2$O$(5_{5,0}-6_{4,3})v2=1$      & 232.687 & 4.00 & $0.19''\times 0.11''$\\\      
D$_2$CO$(4_{2,3}-3_{2,2})$       & 233.650 & 4.41 & $0.19''\times 0.11''$\\    
NaCl$(18-17)$	                 & 234.252 & 4.40 & $0.19''\times 0.11''$\\    
CH$_3$OH(4-5)	                 & 234.683 & 4.65 & $0.19''\times 0.11''$\\     
SO$_2(4_{2,2}-3_{1,3})$	                     & 235.152 & 4.20 & $0.19''\times 0.11''$\\      
HDCO$(6_{2,4}-7_{0,7})$          & 235.887 & 4.46 & $0.18''\times 0.11''$\\    
SO$_2(16_{1,15}-15_{2,14})$	                 & 236.217 & 4.60 & $0.18''\times 0.11''$\\      
SO$(1_2-2_1)$	                 & 236.452 & 4.56 & $0.18''\times 0.10''$\\      
\hline \hline
\end{tabular}
~\\
Notes: $^a$@1km\,s$^{-1}$ spectral resolution
\label{rms}
\end{table}

\begin{figure*}[ht]
	\includegraphics[width=0.99\linewidth]{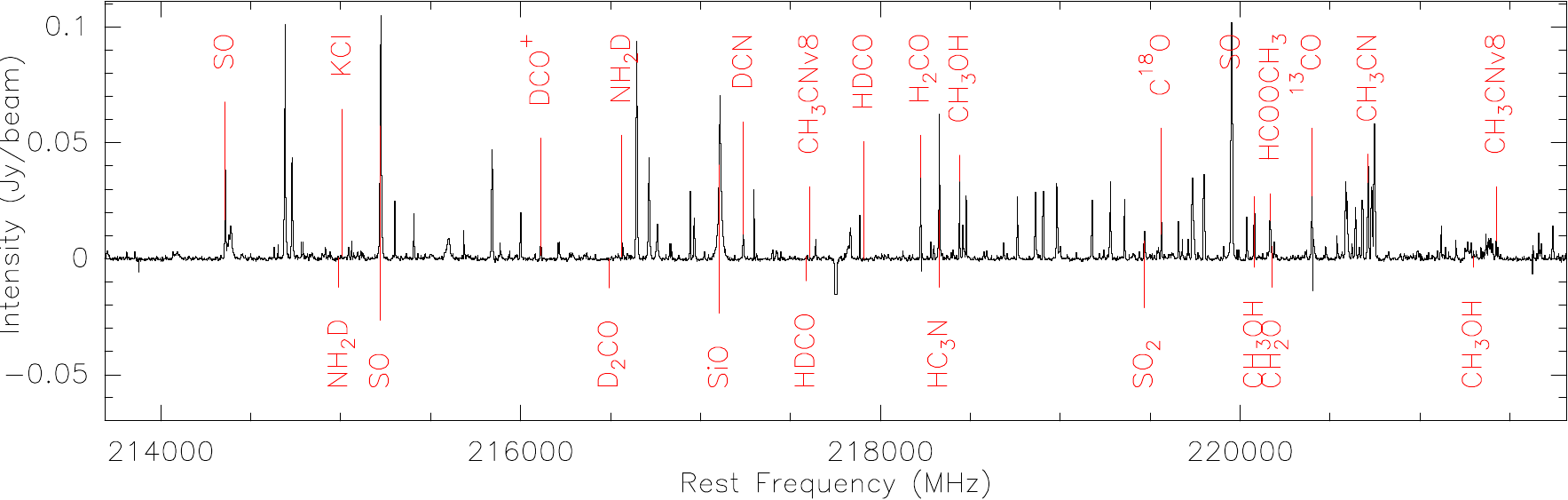}
	\includegraphics[width=0.99\linewidth]{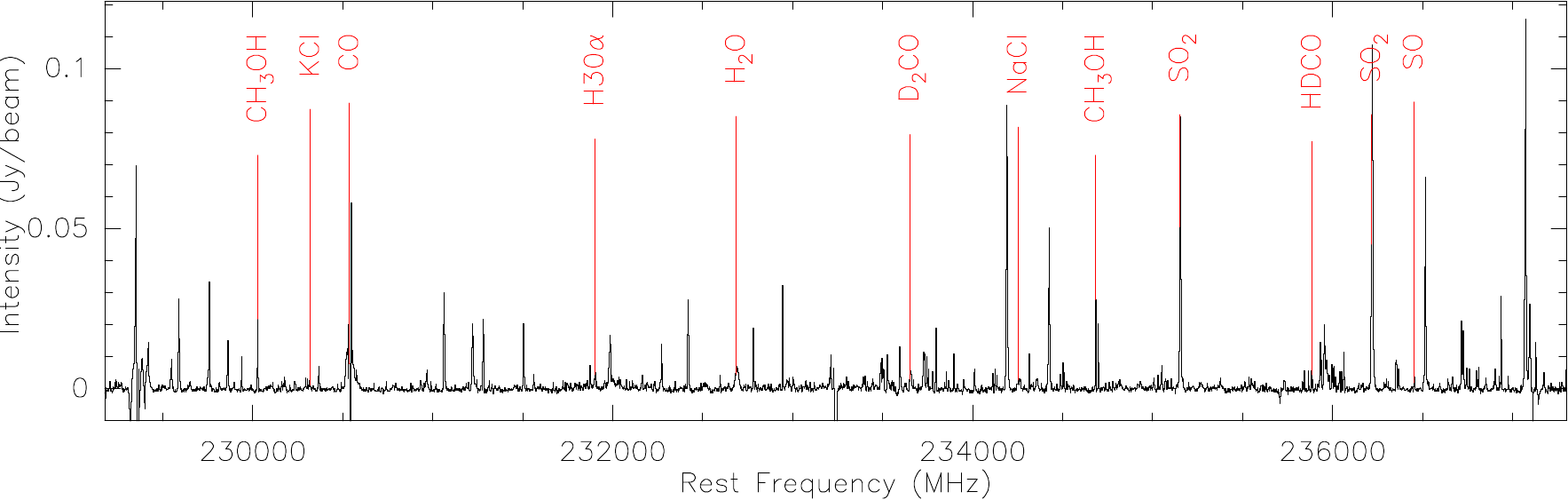}
	\caption{Lower and upper sideband continuum-subtracted average spectra from the central extended disk-like structure with a radius of $0.5''$ from the center in CepA. Lines investigated are labeled (see also Table \ref{rms}).}
	\label{spectra}
\end{figure*}

The new long-baseline capabilities ($\sim$1700\,m) of the Northern Extended Millimeter Array (NOEMA) now allow us to study the properties of high-mass accretion disks in great depth. This study is a high-spatial-resolution follow-up on the successful IRAM large program CORE that investigated the physical and chemical properties of a sample of high-mass star-forming regions on $<$0.1\,pc core-scales (e.g., \citealt{beuther2018b,gieser2021,ahmadi2023}, see also \url{https://www.mpia.de/core}). That project used NOEMA at 1.37\,mm wavelengths in its previous A+B+D configurations (longest baselines at that time $\sim$750\,m) and combined the data with short-spacing information from the IRAM 30\,m telescope. This resulted in high-quality data with exciting results about core fragmentation (e.g., \citealt{beuther2018b,bosco2019,cesaroni2019}), larger-scale disk formation (e.g., \citealt{ahmadi2018,ahmadi2023,moscadelli2021,suri2021}) and the physical and chemical properties of high-mass star-forming regions (e.g., \citealt{gieser2019,gieser2021,gieser2022,mottram2020,beuther2021}). However, the highest spatial resolution of the data was typically around $0.4''$, roughly resolving the rotating toroidal structures on a few 1000\,au scales, but rather insufficient to resolve the embedded disks. Doubling the spatial resolution to $\leq 0.2''$ (a linear resolution of $\leq$140\,au at the distance of CepA of $\sim$0.7\,kpc) with the new NOEMA long baselines shifts high-mass disk investigations to a new level. Within this CORE long-baseline extension program, we now have observed ten regions (CepA, NGC7538IRS1/IRS9/S, AFGL2591, G75.78, IRAS\,21078, IRAS\,23151, IRAS\,23033 and W3H$_2$O, see \citealt{beuther2018b} for source details). While a sample analysis will follow later, several case studies are ongoing (e.g., IRAS\,21078, \citealt{moscadelli2026}), here we presents the results for the closest high-mass disk candidate of the sample, CepA.


\section{Observations}

The data for Cepheus A were obtained with the new long baselines at NOEMA during the winter term 2022/2023 (specifically Feb.~14, 2023) in project W22AL001 (PI H.~Beuther). As a northern hemisphere source, Cep A is ideally suited for long tracks, and it was observed during a 10.5\,h observing run in track-sharing mode with NGC7538IRS9 (the latter region will be discussed in forthcoming work). The covered baseline range was between $\sim$40 and 1663\,m. The phase reference center of the observations was R.A.~(J2000.0) 22$^h$56$^m$17.98$^s$ and Dec.~(J2000.0) 62$^\circ$01$'$49.5$''$. The spectral coverage in the 1.3\,mm band in the lower and upper sideband was between $\sim$213.85 to $\sim$221.60\,GHz and $\sim$229.37 to $\sim$237.17\,GHz, respectively. The spectral resolution of 0.25\,MHz corresponds at 220\,GHz to a nominal velocity resolution of $\sim$0.34\,km\,s$^{-1}$. For our data reduction we used a uniform spectral resolution of 0.5\,km\,s$^{-1}$. While the $v_{\rm lsr}$ of the region is often reported as $-10$\,km\,s$^{-1}$ (e.g., \citealt{moscadelli2009}), we rather consider $\sim -5$\,km\,s$^{-1}$ as the approximate velocity of rest (e.g., \citealt
{sanna2025,nielsen2026} or section \ref{lines}).

Bandpass and flux calibration was conducted with the quasars 3C297 and MWC349, respectively. Gain calibration was performed via regularly interleaved observations of the quasars J0011+707 and J2201+508.

\begin{figure*}[ht]
\includegraphics[width=.99\linewidth,keepaspectratio]{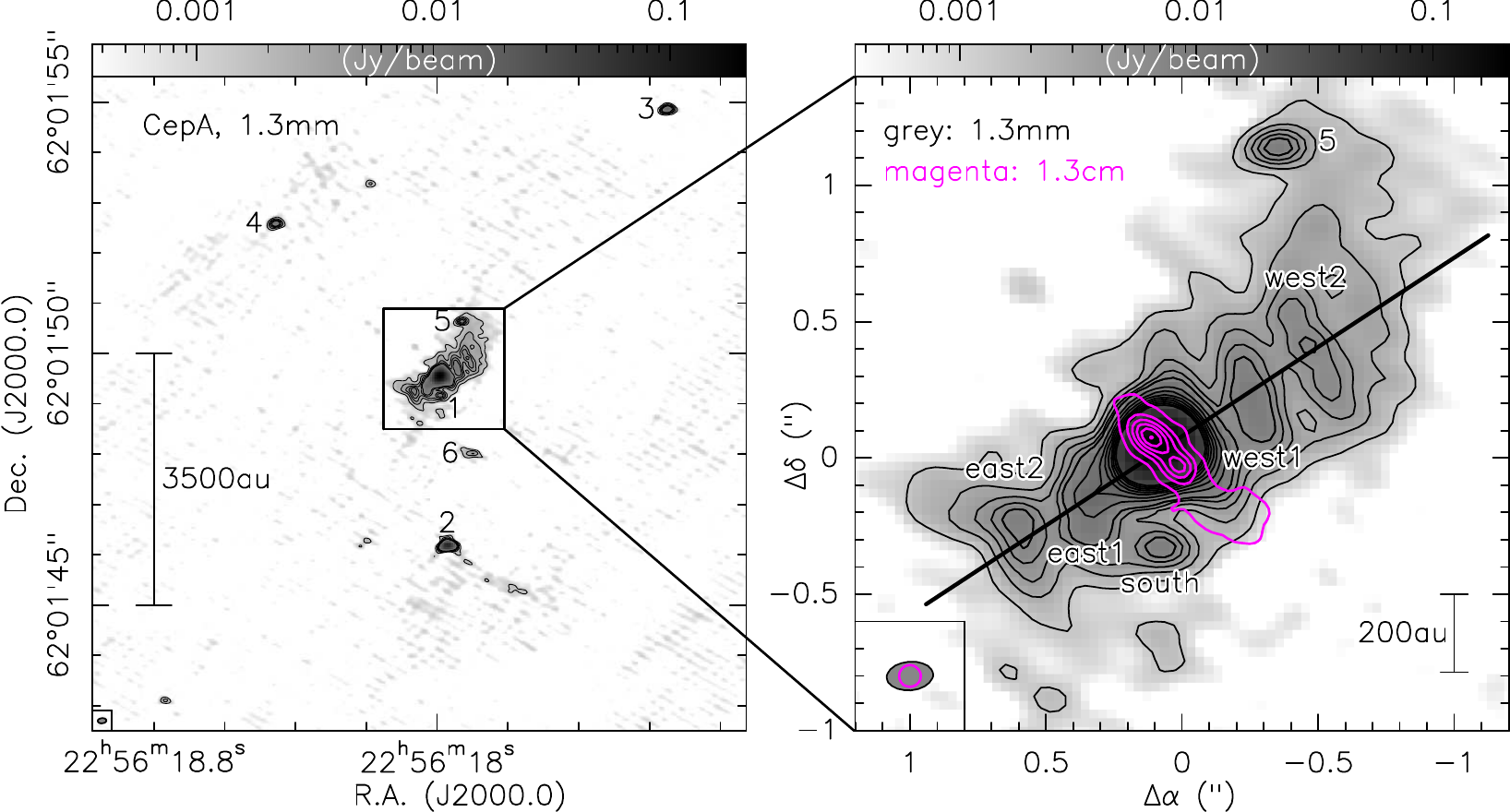}
	\caption{Continuum data for CepA HW2. The left panel shows the large-scale overview of the 1.3\,mm continuum data, and the right panel presents a zoom-in towards the central high-mass disk region. Gray-scale and contours are 1.3\,mm continuum emission starting at $4\sigma$, continuing in $4\sigma$ steps to $20\sigma$, then continuing in $10\sigma$ steps to $100\sigma$. The magenta contours in the right panel outline the ionized jet measured at 1.3\,cm emission by \citet{torrelles1996}. Contour levels start at $4\sigma\approx 0.49$\,mJy\,beam$^{-1}$ and continue in $12\sigma$ steps. Synthesized beams and linear scale-bars are shown in both panels. The thick black line shows the orientation of the intensity and position-velocity cuts in Figs.\,\ref{intensitycut} \& \ref{pv}. Cores and discussed sub-structures are labeled in both panel.}
	\label{cont}
\end{figure*}

Calibration and imaging of the data was conducted within the {\sc GILDAS} framework (\url{http://www.iram.fr/IRAMFR/GILDAS}) with {\sc clic} and {\sc mapping}. The continuum data were created after excluding strong line emission. We combined the lower and upper sideband data to create the continuum dataset for further imaging. The continuum data were self-calibrated within {\sc mapping} in phase-only in three loops with decreasing integration times of 300, 150 and 45\,secs. Imaging was then conducted in uniform weighting to achieve the highest spatial resolution. The final continuum data at a mean frequency of $\sim$225.498\,GHz have a synthesized beam of $0.171''\times 0.105''$ (position angle of $95^\circ$) that corresponds to a geometrical mean of $0.138''$ or 97\,au at 700\,pc distance. The continuum $1\sigma$ rms is 0.36\,mJy\,beam$^{-1}$.

While the self-calibration improved the continuum data significantly with an increase of the peak signal-to-noise ratio from 126 to 586, no clear improvement of the line data was recognizable after applying the solutions from the self-calibration. Therefore,  for the spectral line data, we used the non-self-calibrated data products. While Fig.~\ref{spectra} shows an overview of the entire lower and upper sideband averaged over the central extended disk-like structure, for the analysis in this paper we focus on only a few spectral lines (also marked in Fig.~\ref{spectra}). The rms in 0.5\,km\,s$^{-1}$ channels is typically a few mJy\,beam$^{-1}$ with a synthesized beam comparable to the continuum data as described above. Table \ref{rms} lists the actual rms and synthesized beam values for the explored spectral lines.

\section{Results}

\subsection{Continuum emission}

The 1.3\,mm dust continuum map shown in Fig.~\ref{cont} identifies at least six structures within the field of view. While the central elongated structure is the disk candidate (e.g., \citealt{patel2005,comito2007,sanna2025}) and our main region of interest, there are at minimum five more cores in the close environment. Focusing on the central elongated structure (Fig.~\ref{cont} right panel) it is not smooth, but in addition to the central peak one finds several more peaks within it (west1, west2, east2), plus a south-eastern elongation from the peak (east1) and additional separate peaks offset from the assumed disk mid-plane, towards the north-west (core \#5) and south of the main peak (south). The central peak is also the location of the well-known cm jet with northeast-southwest orientation (magenta contours in Fig.~\ref{cont} right panel).

Comparing this new highest-resolution mm continuum map with previous studies of the region, we clearly identify far more sub-structures. All previously existing (sub)mm maps showed only an elongated structure in the northwest-southeastern direction but no more sub-features \citep{patel2005,comito2007,beuther2018b}. This elongated northwest-southeast structure was so far considered as the disk driving the central jet (e.g., \citealt{patel2005}).

\subsection{Structures, masses and column densities} 
\label{mass_col}

Before studying the proposed disk in more detail, we estimate gas column densities $N$ and masses $M$ for all structures in the region. We first identified cores with the clumpfind approach \citep{williams1994} which is adequate for the clearly identifiable substructures visible in the 1.3\,mm continuum map (Fig.~\ref{cont}). We only identified structures as cores if detected above $10\sigma$, and fluxes were extracted then above $5\sigma$. While the clumpfind algorithm divides the central extended structure into several sub-cores, based on the coherent velocity structure discussed below (Sect.~\ref{lines}), we concatenated them into one single extended disk-like structure and labeled that as core \#1. This source \#1 has further sub-structures that will be discussed in more detail below, in particular the extensions to the east and west (east1, east2, west1, west2) as well as an additional peak $\sim$0.4$''$ south of the main peak. All identified cores and sub-structures are labeled in Fig.~\ref{cont}. Peak fluxes and integrated flux densities are listed in Table \ref{masses}.

To use these core parameters for estimates of the column densities $N$ and the masses $M$, we assumed optically thin dust emission following \citet{hildebrand1983} and \citet{schuller2009} (see also Section \ref{disk} and Fig.~\ref{tau}). We assumed a gas-to-dust mass ratio of 150 \citep{draine2011} and a dust opacity $\kappa=0.9$\,cm$^2$g$^{-1}$ from \citet{ossenkopf1994} for grains with thin ice mantles at densities of $10^6$\,cm$^{-3}$.
Regarding the temperatures, for the extended disk-like structure \#1 we used an approximate mean temperature of 225\,K from our CH$_3$CN spectral line fitting (see Section \ref{lines}), assuming that gas and dust are well coupled. For the other sources, we use a lower temperature of 25\,K. The derived core masses $M$ and column densities $N$ are presented in Table \ref{masses}.

\begin{table}[htb]
\centering
\caption{Core parameters}
\begin{tabular}{lrrrr}
\hline \hline
\# & $S_{\rm peak}$ & $S_{\rm int}$ & $N^a$ & $M^a$ \\
   & $\left(\frac{\rm Jy}{\rm beam}\right)$ & (Jy) & ($10^{25}$cm$^{-2}$) & (M$_{\odot}$) \\
   \hline
1$^b$ & 0.210 & 0.662 & 4.6  & 0.75 \\
2$^c$ & 0.074 & 0.096 & 17.7 & 1.19 \\
3 & 0.014 & 0.018 & 3.4  & 0.23 \\
4 & 0.011 & 0.014 & 2.7  & 0.18 \\
5 & 0.009 & 0.013 & 2.1  & 0.17 \\
6 & 0.004 & 0.006 & 1.1  & 0.07 \\
\hline \hline
\end{tabular}
~\\
$^a$Lower limits because of missing flux.\\
$^b$HW2 in \citet{torrelles1998}\\
$^c$HW3c in \citet{torrelles1998}
\label{masses}
\end{table}

Comparing our new results with those from the earlier original CORE studies at a lower angular resolution of $0.44''\times 0.38''$, it is interesting to note that in the original CORE data only cores \#1 and \#2 were identified \citep{beuther2018b}. This is on the one hand because of the conservative clumpfind approach where only sources $>10\sigma$ were reported (which excluded cores \#3 and \#4), and on the other hand sources \#5 and \#6 were not individually identified but remained part of the main central structure \#1. While the measured flux differences for the very compact core \#2 between the new and old data is less than 10\%, we find that for the more extended disk-like structure \#1 our new long-baseline data recover only about 50\% of the flux found in the original CORE study. Hence, missing flux is a severe issue in determining the parameters in particular for the central disk-structure \#1. In addition to that, the central peak of the elongated disk structure shows high optical depth (Fig.~\ref{tau}). Therefore, the measured masses typically $\leq 1.2\,M_{\odot}$ are clearly only lower limits to the actual core masses. Nevertheless, the peak fluxes reflect the actual column densities better, and peak column densities are typically in excess of $10^{25}$\,cm$^{-2}$.

\begin{figure}[ht]
	\includegraphics[width=0.99\linewidth ,keepaspectratio]{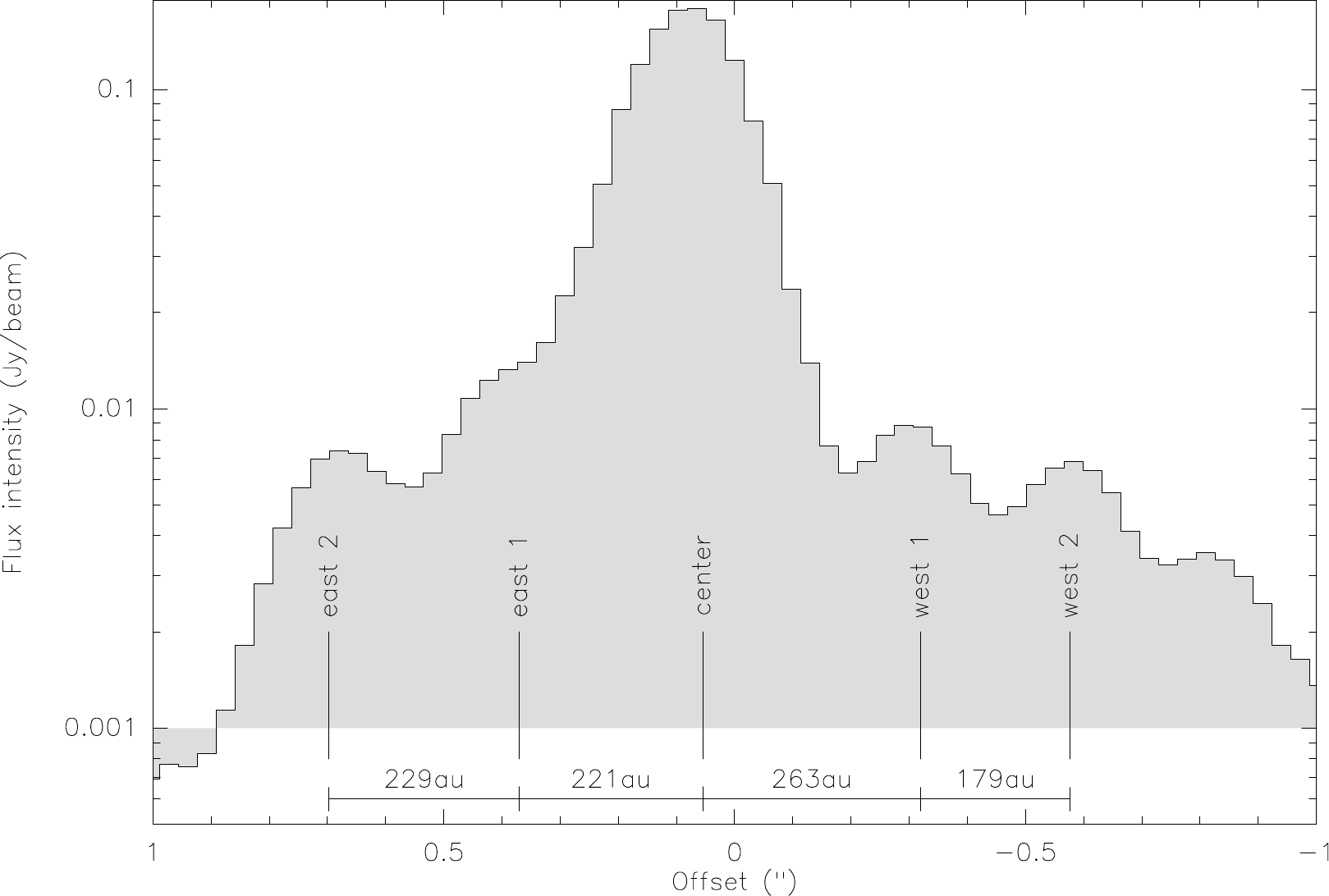}
	\caption{Intensity cut along the 1.3\,mm continuum map as outlined in the right panel of Fig.\,\ref{cont}. The main continuum peaks and their separations are marked.}
	\label{intensitycut}
\end{figure}

\begin{figure}[ht]
	\includegraphics[width=0.99\linewidth ,keepaspectratio]{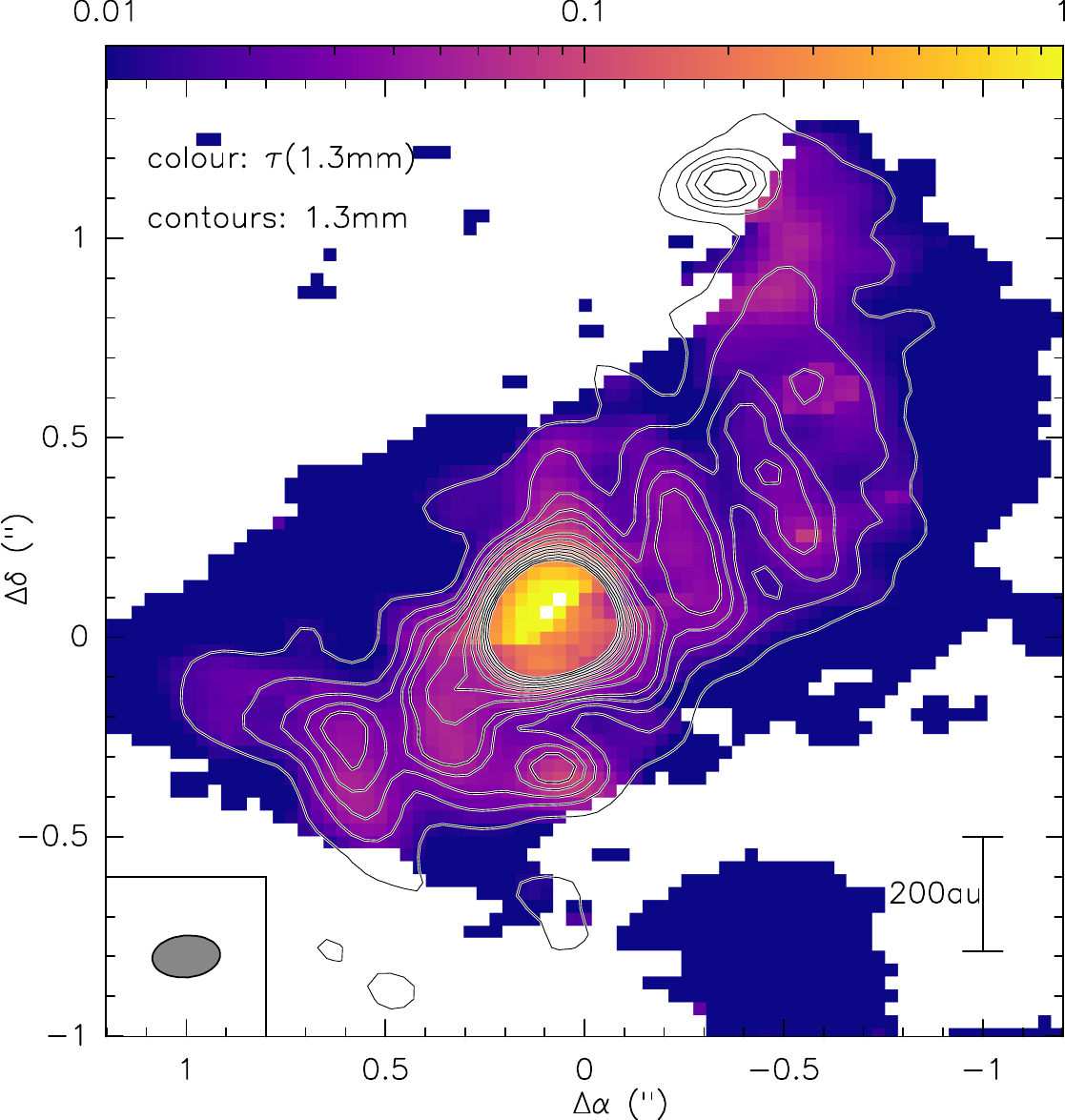}
	\caption{Optical depth $\tau_{\nu}$ map at 1.3\,mm. The color scale shows the $\tau_{\nu}$ map and the contours outline the 1.3\,mm continuum emission in $4\sigma$ steps.}
	\label{tau}
\end{figure}

\subsection{The central disk-like structure}
\label{disk}

We now focus on the proposed central disk structure labeled here as core \#1. To look at the intensities in more detail, Fig.~\ref{intensitycut} presents an intensity cut along its major axis as outlined in the right panel of Fig.~\ref{cont}. The entire extent of the structure is $1.9''$ or 1330\,au at the given distance of 0.7\,kpc. Figures \ref{cont} and \ref{intensitycut} both show, in addition to the central peak, at least three more peaks. The central peak exhibits an elongation towards the southeast that we consider as a fourth central structure. Figure \ref{intensitycut} marks the positions of these four additional peaks as east1, east2 and west1, west2. The separations between these emission peaks appears comparably symmetric towards the northwest and southeast with projected separations between 179 and 263\,au (Fig.~\ref{intensitycut}).

\begin{figure*}[ht]
	\includegraphics[width=0.99\linewidth ,keepaspectratio]{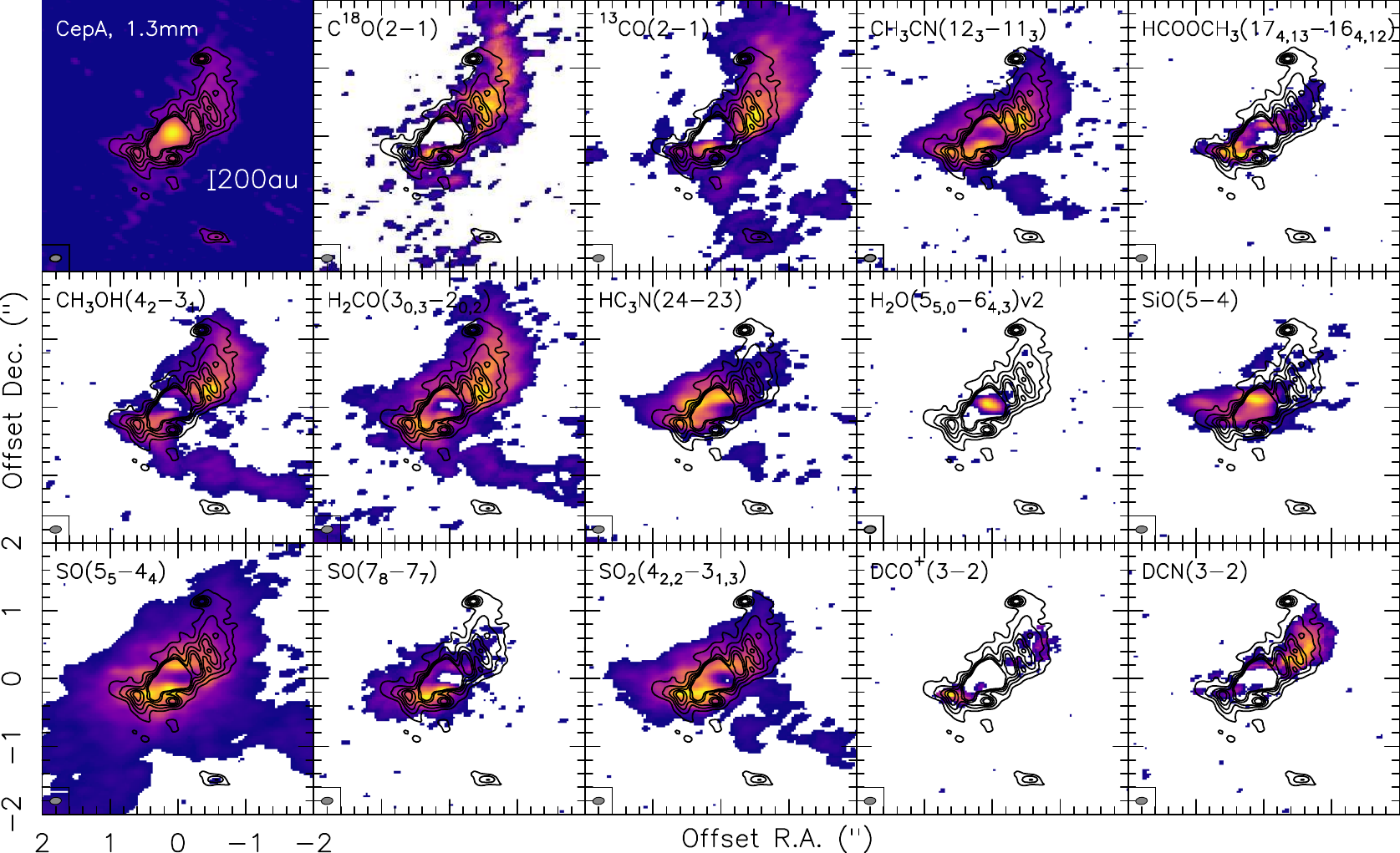}
	\caption{Integrated intensity maps for selected molecules (integration range typically [$-13$,$-1$]\,km\,s$^{-1}$, only for CH$_3$CN, HCOOCH$_3$, and H$_2$O slightly different [$-15$,3], [$-13$,3], [$-30$,7], Table \ref{rms}). The top-left panel shows in color and contours the 1.3\,mm continuum emission, whereas in all other panels the color presents the moment 0 maps created above the $5\sigma$ values, and the contours again the 1.3\,mm continuum. The contour levels are always in $4\sigma$ steps. The synthesized beam is shown in all panels, the the top-left panel also presents a scale-bar.}
	\label{mom0}
\end{figure*}

The flattened continuum structure perpendicular to the cm jet indicates that the disk should be roughly edge-on. While earlier studies prefer less inclined configurations (e.g., 62$^o$ from lower resolution data, \citealt{patel2005}, or $51\pm 11^o$ indirectly from light echo studies, \citealt{stecklum2025}), our data and later discussion (Sec.~\ref{disk-discussion}) favor a more closer to edge-on geometry. In this configuration, the question is whether the intensity peaks towards the east and west are real structures within the disk, for example, protostellar condensations  (e.g., \citealt{ahmadi2019}), or rings akin to those observed in many low-mass disks (e.g., \citealt{andrews2018}), or whether they are potentially caused by opacity effects. We estimate the optical depth of the 1.3\,mm continuum emission following Appendix A.3 in \citet{frau2010}:
\begin{eqnarray}
    \tau_{\nu}=-{\rm ln}\left(1-\frac{S_{\nu}}{\Omega B_{\nu}(T)}\right)
  \end{eqnarray}
with the flux density $S_{\nu}$, the synthesized beam $\Omega$ and the Planck function $B_{\nu}(T)$ depending on the temperature $T$. The latter is inferred from the CH$_3$CN emission as outlined below in section \ref{lines}. Figure \ref{tau} presents the derived optical depth map of the CepA disk-like region. While the optical depth towards the central peak is greater than 1, all extended emission also in the secondary peaks towards the east and west have comparably low optical depth around 0.05 or even lower. Hence, the 1.3\,mm continuum emission indeed traces the significant structures along the line of sight. We will get back to the potential sub-structure of the CepA disk in Section \ref{disk-discussion}.

\subsection{Spectral line emission}
\label{lines}

As shown in Figure \ref{spectra},  the region is very line rich, and our broad spectral bandpass of almost 16\,GHz covers a plethora of spectral lines. In the context of this paper, we concentrated on those outlined in Table \ref{rms}. These encompass rather standard lines like CO and its isotopologues, and then concentrate on dense gas tracers (e.g., CH$_3$CN, HCOOCH$_3$, CH$_3$OH), shock tracers (e.g., SiO, SO, SO$_2$), deuterated species (e.g., DCO$^+$, DCN, D$_2$CO), salt and water lines recently suggested as disk tracers (e.g., KCl, H$_2$O, \citealt{ginsburg2023,yang2026}), and the recombination line H30$\alpha$ to trace the ionized gas. Since we present here data from only the new extended configuration, one should always keep in mind that missing flux affects the extended emission for any tracer. Combining these data with the existing shorter baseline observations (e.g., \citealt{gieser2021}) is left for future studies.

Figure \ref{mom0} presents integrated intensity maps of example species and lines. Emission dips towards the main continuum peak are typically caused by the high optical depth of the continuum and associated absorption toward that (see also Fig.~\ref{spectra_ch3cn}, 2nd panel for an example spectrum). Exceptions of that are some deuterated species like DCO$^+$ which show no emission towards the central continuum peak. This is likely associated with the highest temperatures there, and DCO$^+$ being known to chemically only emit in cold gas (e.g., \citealt{pety2007,parise2009,beuther2022b}). In general, most of the molecular line emission is found towards the continuum emission, for some lines also beyond that. Just the H$_2$O line emission is particularly compact and only found towards the central peak position. We point out that the salt lines of KCl and NaCl are not detected towards CepA (for comparison see \citealt{ginsburg2023} or \citealt{yang2026}). The recombination line H30$\alpha$ is only detected weakly at the south-eastern edge of the disk-like structure but not toward the central jet, differently to what has recently been reported for the jet of W75N(B)-VLA3 \citep{sanchez-monge2025}.

\begin{figure*}[ht]
	\includegraphics[width=0.99\linewidth ,keepaspectratio]{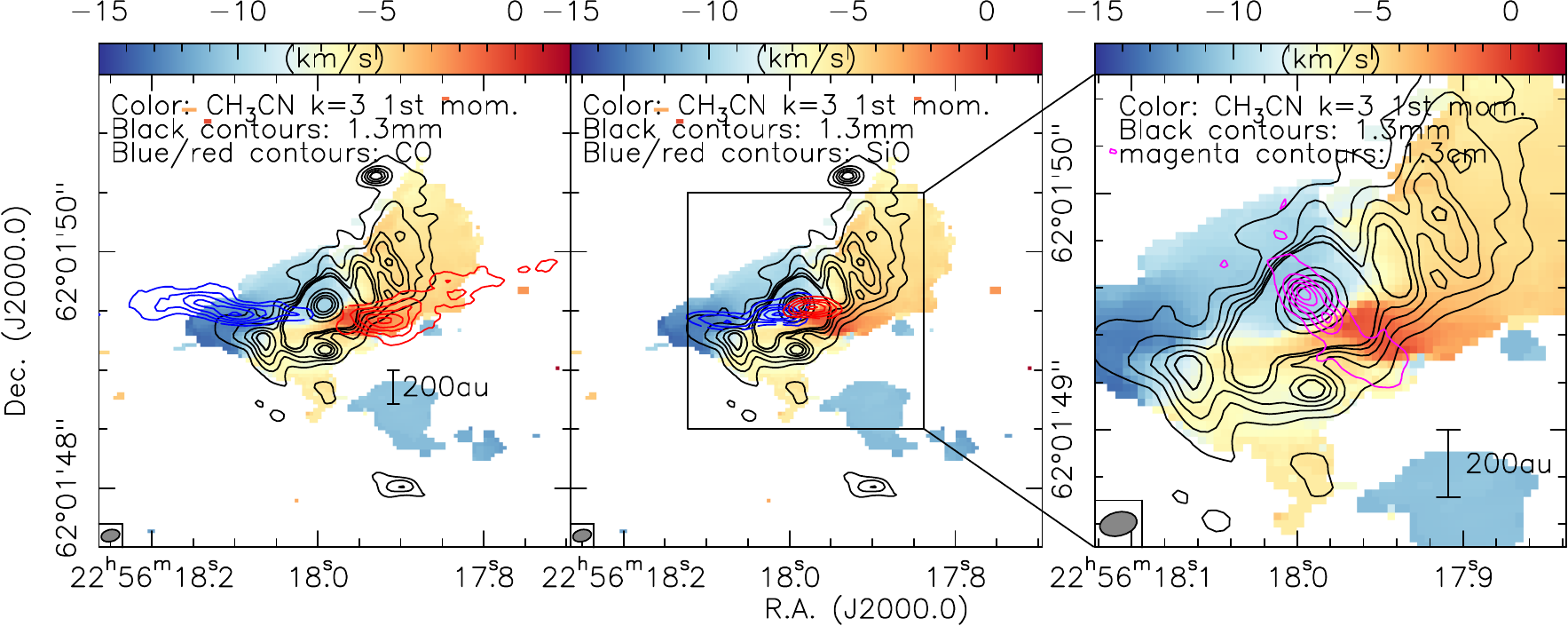}
	\caption{Overlay of CH$_3$CN, 1.3\,mm and outflow data. All three panels show in color and contours the CH$_3$CN$(12_3-11_3)$ first moment and 1.3\,mm continuum maps. Contour levels are from 4$\sigma$ to $20\sigma$ in $4\sigma$ steps and then continue in 100$\sigma$ steps. The blue and red contours in the left and middle panel outline the east-west outflow in CO(2--1) (blue [-50,-30]\,km\,s$^{-1}$, red [20,40]\,km\,s$^{-1}$) and SiO(5--4) (blue [-50,-30]\,km\,s$^{-1}$, red [10,30]\,km\,s$^{-1}$), respectively. The right panel shows the 1.3\,cm continuum jet \citep{torrelles1996} in magenta contours. Synthesized beams and scale-bars are presented as well.}
	\label{overlay}
\end{figure*}

\begin{figure*}[ht]
\sidecaption
    \includegraphics[width=12cm,keepaspectratio]{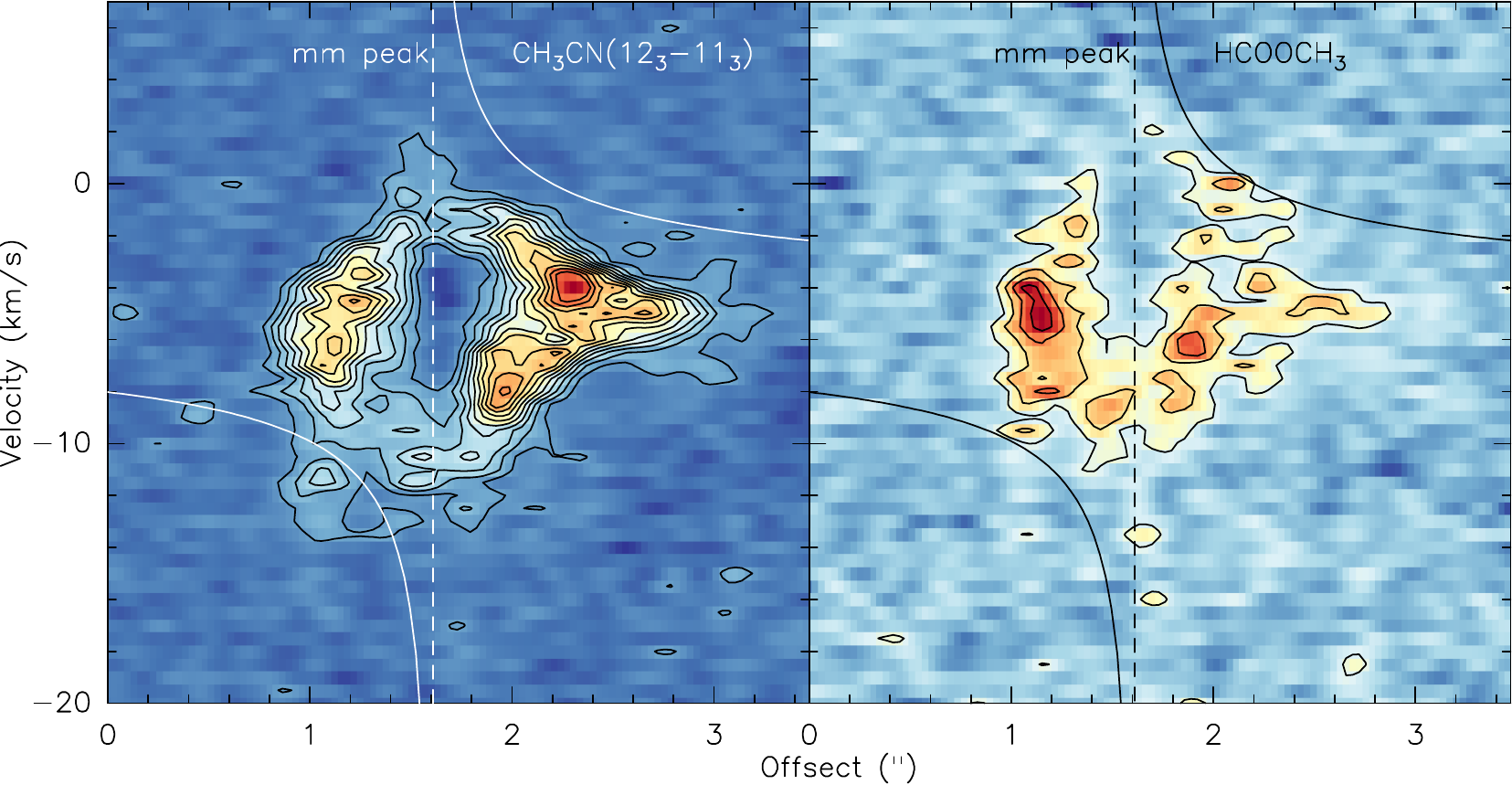}
	\caption{Position-velocity cuts along the line shown in Fig.~\ref{cont} (right panel, offset 0 is south-east). The left and right panels show the emission from the CH$_3$CN$(12_3-11_3)$ and HCOOCH$_3(17_{4,13}-16_{4,12})$ lines, respectively. Contours are in both panels in $3\sigma$ levels. The dashed line marks the central position, and the curves correspond to a Keplerian profile around a 12\,M$_{\odot}$ central object (no fit to the data). A disk inclination angle of 80\,deg is assumed.}
	\label{pv}
\end{figure*}

\subsubsection{Disk and outflow kinematics}

To Investigate the kinematics of the proposed disk and the associated outflows, Figure \ref{overlay} presents the velocity structure in the CH$_3$CN$(12_3-11_3)$ line in color-scale and the blue- and red-shifted emission of the outflow tracers CO(2--1) and SiO(5--4) in contours. For comparison, Fig.~\ref{mom1} shows the 1st moment maps (intensity-weighted peak velocities) for all molecules where the integrated emission is shown in Fig.~\ref{mom0}. While in the past, mainly the ionized northeast-southwest jet was reported (magenta contours in the right panel of Fig.~\ref{overlay}), the CO and SiO high-velocity data (up to $\pm 45$\,km\,s$^{-1}$ from the $v_{\rm lsr}$) clearly show a second east-west outflow in the molecular line emission. This high-velocity central east-west molecular outflow may correspond to the larger-scale CO and H$_2$ outflows \citep{rodriguez1980,cunningham2009} and was  tentatively suggested also in \citet{comito2007}. In addition to that, \citet{jimenez2007} detected a western 7\,mm continuum extension from the central jet that may also be related to this east-west outflow. Both, the ionized northeast-southwest jet and the molecular east-west outflow appear to stem from the central peak position. Having two outflows/jets from one emission peak at a spatial resolution of $\sim$97\,au indicates that this innermost region should host an unresolved binary already.

Focusing on the elongated disk structure, the velocity structure extracted from the CH$_3$CN$(12_3-11_3)$ is interesting. While one sees an overall shift from red- to blue-shifted from the northwest to the southeast, perpendicular to the ionized jet, there is significant velocity substructure. In particular, close to the central continuum emission peak, the red-blue-shifted structures are elongated somewhere inbetween the west-east molecular outflow (identified in the CO and SiO emission) and the northeast-southwest ionized jet. This indicates, while CH$_3$CN is often regarded as a good disk tracer, its velocity structure can also be affected by the innermost jets and outflows (see also \citealt{leurini2011,moscadelli2013,beltran2018,busch2026}). That needs to be kept in mind for further interpretation. For comparison, we also investigated the very compact H$_2$O emission where integrated emission, 1st and 2nd moments are presented in Fig.~\ref{overlay_h2o}. In its compact structure, the velocity gradient is more north-south and the velocity dispersion in the 2nd moment map exhibits line widths significantly in excess of 10\,km\,s$^{-1}$. This can also be interpreted in the framework that the H$_2$O emission largely traces outflow emission, probably affected by the ionized northeast-southwest jet and the east-west molecular outflow.

To look in more detail at the dense gas velocity structure, we created position-velocity (pv) cuts along the main disk-like structure as outlined in Fig.~\ref{cont}. These pv cuts for CH$_3$CN and another dense gas tracer HCOOCH$_3$ are presented in Fig.~\ref{pv}. While HCOOCH$_3$ exhibits a lower signal-to-noise ratio, the overall structures are similar. We also show what profile a Keplerian disk around a 12\,M$_{\odot}$ central object would show but we stress that this is no fit to the data. For the Keplerian curve we assume a disk inclination angle of 80\,deg, close to an edge-on configuration. The 12\,M$_{\odot}$ for a central object was estimated by \citet{debuizer2017} from SOFIA spectral energy distribution fits and is similar to a B1 star of 13\,M$_{\odot}$ \citep{lang1992}, corresponding to the total luminosity of the region of $1.5\times10^4$\,L$_{\odot}$ \citep{beuther2018b}. While the Keplerian curves resemble some of the pv-structures, the highest-velocity gas towards the center is barely detected. And even more important, there is significant emission in the other two quadrants of these plots which may be attributed for example to infalling gas (e.g., \citealt{ohashi1996,ahmadi2019,williams2022}). Hence, these pv-cuts do not resemble well typical Keplerian velocity profiles. \citet{sanna2025} recently also reported sub-Keplerian velocity profiles and gas infall down to radii of 250\,au.

\begin{figure}[ht]
	\includegraphics[width=0.99\linewidth ,keepaspectratio]{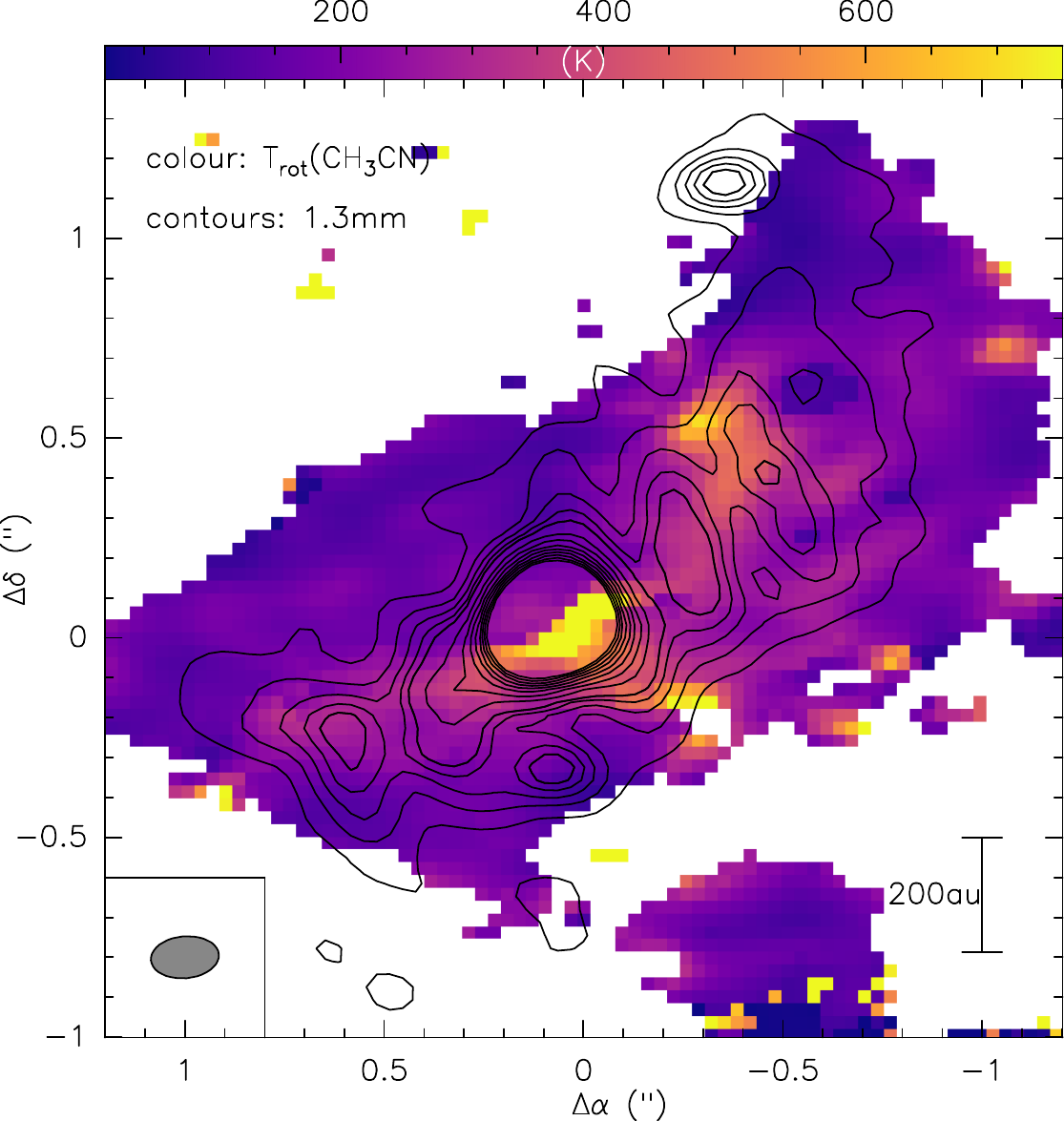}
	\caption{Rotation temperature map from CH$_3$CN$(12_k-11_k)$. The color scale shows the temperature map and the contours outline the 1.3\,mm continuum emission in $4\sigma$ steps. A synthesized beam and linear scale bar are shown as well. }
	\label{trot}
\end{figure}

\subsubsection{Temperature analysis}

In addition to the kinematic analysis, the CH$_3$CN$(12_k-11_k)$ data with $k=0...7$ allow us to derive rotational temperatures pixel by pixel for the entire central disk region where CH$_3$CN is well detected. At the given high densities, the rotational temperatures, that describe the level population in a Boltzmann distribution, should correspond to the actual gas temperatures in the disk layers CH$_3$CN is sensitive to. The latter does not only depend on the densities but also on the optical depth. One clearly sees that towards high column density positions, the CH$_3$CN is strongly suppressed which indicates high optical depth and self absorption (e.g., second panel in Fig.~\ref{spectra_ch3cn}).

For the fitting of the entire CH$_3$CN map we employed the package XCLASS \citep{moeller2017}. XCLASS models the line emission in local thermodynamic equilibrium (LTE) for an isothermal homogeneous object using the VAMDC and CDMS databases\footnote{\url{http://www.vamdc.org}, \url{https://cdms.astro.uni-koeln.de}} for spectral lines \citep{mueller2001}. In the frequency range from 220.524 to 220.762\,GHz we fitted simultaneously the CH$_3$CN$(12_k-11_k)$ (with $k=0...7$) and CH$_3^{13}$CN$(12_k-11_k)$ (with $k=0...5$) emission. The resulting CH$_3$CN temperature map is presented in Fig.~\ref{trot}. While the extremely high temperatures towards the central position are unreliable because of the mentioned high optical depths and corresponding strong self-absorption (Fig.~\ref{spectra_ch3cn}, 2nd panel), the remaining part of the temperature map is reasonable for such a hot core region with an approximate mean temperature of $\sim$230\,K. We stress that this should not be a disk midplane temperature but rather higher disk layers that are more exposed to the luminosity of the central region.

\begin{figure*}
\sidecaption
    \includegraphics[width=12cm,keepaspectratio]{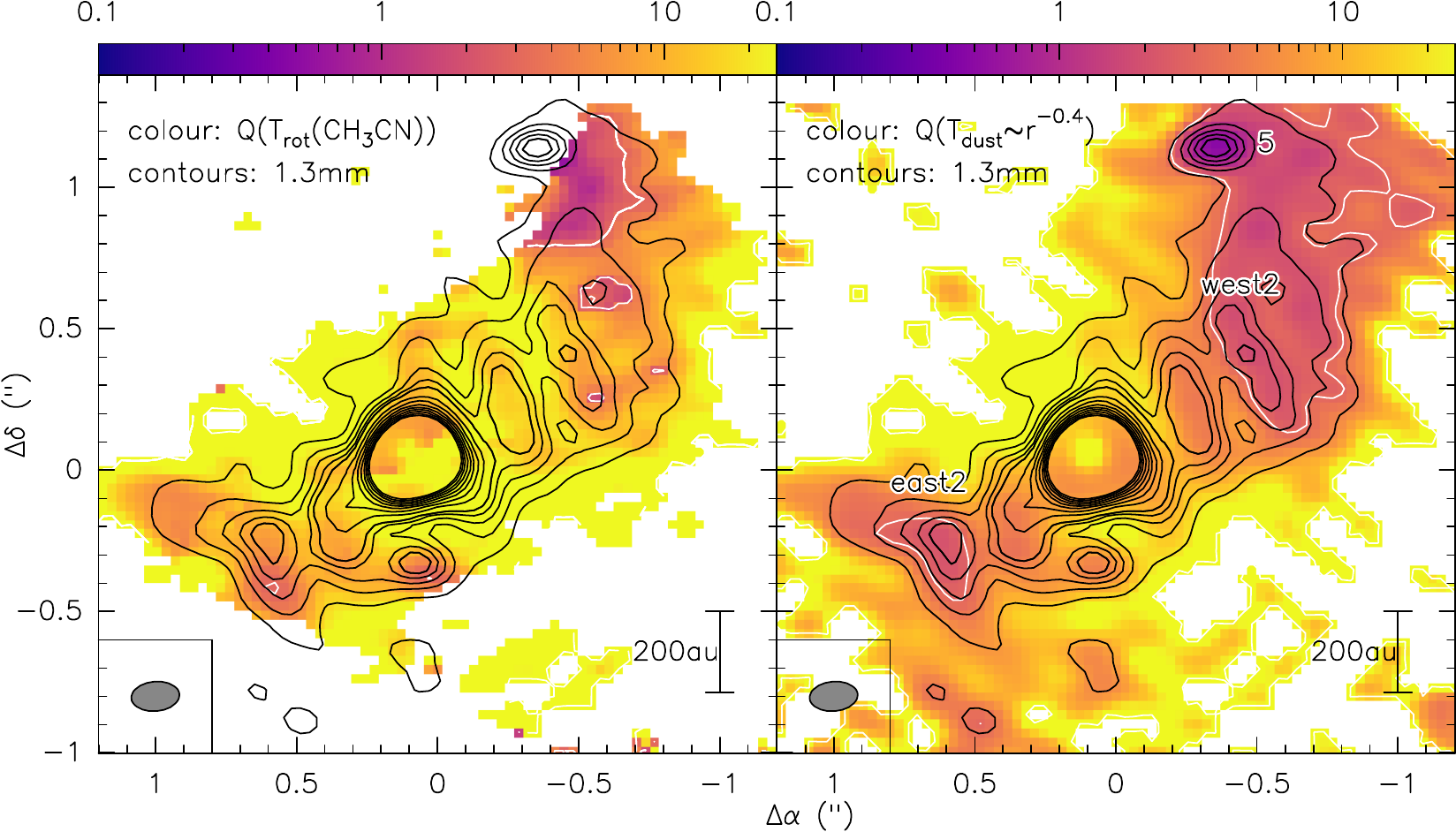}
	\caption{Toomre $Q$ maps for the CepA disk. The left panel shows the Toomre $Q$ map using the CH$_3$CN rotational temperature map as input, whereas the right panel presents the estimated Toomre $Q$ map assuming a temperature gradient $T\propto r^{-0.4}$. The white contour shows the $Q=3$ line. The black contours outline the 1.3\,mm continuum emission in $4\sigma$ steps. Synthesized beam and linear scale bars are shown in both panels. The outermost eastern and western dust continuum positions and source \#5 are marked in the right panel.}
	\label{toomre_q}
\end{figure*}

\subsubsection{Stability analysis}

To investigate the disk stability more, we aim to derive a Toomre $Q$ map of the central disk-like region. While the Toomre $Q$ analysis was originally derived for disks of stars \citep{toomre1964}, it has since then be used in many environments for stability analysis (e.g., \citealt{binney2008}), in particular also for accretion disks around forming stars (e.g., \citealt{gammie2001,kratter2010,klassen2016,ahmadi2023}). The Toomre $Q$ parameter is defined as:
    \begin{eqnarray}
    Q=\frac{c_s\kappa}{\pi G\Sigma} \label{Q}
\end{eqnarray}
with the sound speed $c_s$, the epicyclic frequency $\kappa$, the gravitational constant G and the surface density $\Sigma$. For a Keplerian disk, the epicyclic frequency $\kappa$ corresponds to the Keplerian angular velocity $\Omega$ (here estimated for a 12\,M$_{\odot}$ central object). In general, for thin disks axisymmetric instabilities can occur for $Q\leq 1$ (e.g., \citealt{gammie2001}), but for disks with finite scale height, disks can also be stable for slightly lower $Q$ values down to $\sim$0.6 (e.g., \citealt{kim2007,behrendt2015}). In contrast to that, non-axisymmetric instabilities like spiral arms can also occur at slight larger $Q$ values up to $\sim$2 (e.g., \citealt{binney2008}).

One important input parameter for the $Q$ estimate is the temperature of the disk. This is needed for the sound speed $c_s$ and the surface density. As a first approach, we use as temperature the rotational temperature derived before from the CH$_3$CN emission (Fig.~\ref{trot}). In addition to that, to estimate the epicyclic frequency $\kappa$ we again assume a close-to edge-on disk with an inclination angle of 80\,deg. The resulting Toomre $Q$ map is shown in the left panel of Fig.~\ref{toomre_q}. Within this Toomre $Q$ map one finds on average comparably large $Q$, mostly larger than 2, and often also larger than 10. Only towards the north-western edge of the map, the Toomre $Q$ values fall below 2. This analysis would imply a large stable disk. This in itself already appears unlikely since this disk-like structure shows several significant emission peaks indicative of fragmentation (Figs.~\ref{cont} \& \ref{intensitycut}).

However, as outlined before, the CH$_3$CN temperature map does not trace the disk midplane where the instabilities should occur but the CH$_3$CN emission rather traces higher surface layers of the disk (see also \citealt{motte2025} for differences between molecular line based gas temperatures and dust temperatures). Hence, it is unlikely to be a good proxy for the midplane disk structure. Since we do not have a better disk midplane tracer, we can try to estimate the temperature from classical power-law temperature distributions for protostellar accretion disks. Assuming a dust sublimation temperature of 1400\,K at 5\,au in a disk irradiated by a central $1.5\times 10^4$\,L$_{\odot}$ source, the temperature can then fall like $T\propto r^{-\alpha}$. While classical disk solutions get $\alpha=0.5$ (e.g., \citealt{kenyon1987}), radiative transfer models and observations for embedded disks also find lower values like $\alpha=0.4$ (e.g., \citealt{whitney2003,gieser2021,gieser2023}). We are using the latter value in the following. With that dust temperature profile, we can derive a new Toomre $Q$ map as presented in the right panel of Fig.~\ref{toomre_q}. While the values towards the central peak are still large in excess of 2, the Toomre $Q$ values significantly drop going further outside. Toward the outer peaks east2 and west2, $Q$ is typically around 2 or even lower. Towards the north-western source \#5, Toomre $Q$ becomes lowest around 0.5, clearly in the unstable regime. 

Other uncertainties affect the $Q$ determination too. As outlined in Section \ref{mass_col}, missing flux is important for the column density estimates, making them lower limits. Furthermore, the disk is not necessarily everywhere in Keplerian rotation, and \citet{sanna2025} reported sub-Keplerian velocity structures for the outer disk. In a sub-Keplerian configuration, the angular velocity $\Omega$ would be lower. Combining the lower limit on $\Sigma$ with the upper limit on $\Omega$ in Eq.~\ref{Q} implies that the derived Toomre $Q$ values should be considered as upper limits.

While the absolute Toomre $Q$ values should be considered with caution because of the large uncertainties especially in the temperature structure, one nevertheless finds decreasing $Q$ values towards the fragments within the main disk-like structure CepA. \citet{oliva2020} discuss that for a fragmented disk the disk gas can still have high $Q$ values and only the fragments themselves exhibit very low $Q$. Hence, a general scenario of disk fragmentation appears plausible (see Section \ref{disk-discussion}).

\section{Discussion}

\subsection{Multiplicity and multiple outflows}

The presence of an ionized jet in CepA has been long established (e.g., \citealt{torrelles1996,gomez1999}, Fig.~\ref{cont}), and the perpendicular orientation of dust continuum and dense gas line emission has hence been considered as disk structure (e.g., \citealt{comito2007,beuther2018b,ahmadi2023,sanna2025}). There has long been the suggestion of high multiplicity in this region, with up to six potential protostellar objects discussed within the central $\sim 1''$ \citep{curiel2002,martin-pintado2005,comito2007}. This is consistent with the previous suggestion that there may also be multiple outflows in this region \citep{comito2007}.

\begin{figure*}
\sidecaption
\includegraphics[width=12cm,keepaspectratio]{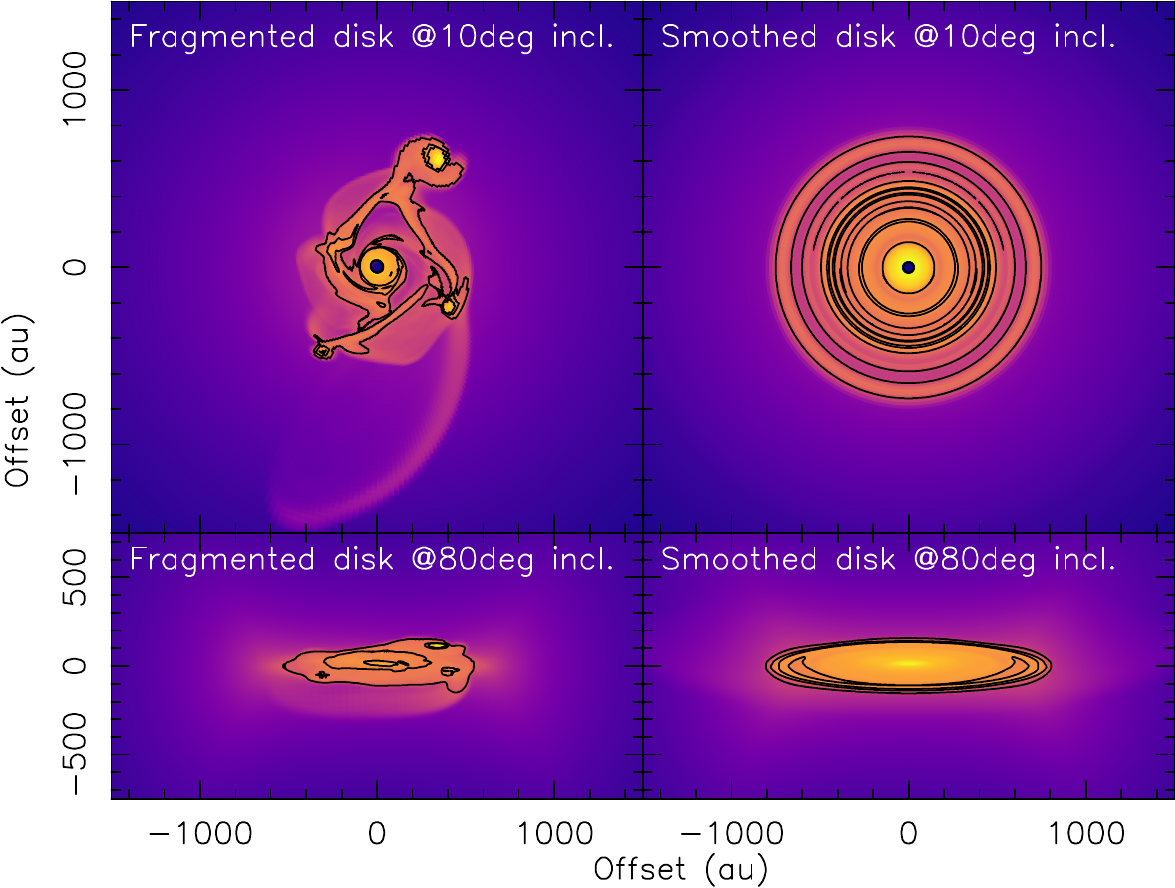}
	\caption{Post-processed 1.37\,mm continuum images based on \citet{ahmadi2019}. The left panels show the fragmented disk at 10 and 80\,deg inclination (top and bottom, respectively, similar to Fig.~3 in \citealt{ahmadi2019}). The right panel uses the same data, just smoothing out the emission azimuthally to create ring-like structures, again top and bottom 10 and 80\,deg inclination. Contour levels are from 10/20/30\% of the peak emission.}
	\label{ahmadi}
\end{figure*}

Our new high-resolution data now clearly confirm the presence of a second east-west outflow in high-velocity CO and SiO emission (Fig.~\ref{overlay}) that is potentially related to the large-scale east-west outflows observed in CO and H$_2$ emission \citep{rodriguez1980,cunningham2009}. The origin of this second outflow appears to be the central peak position, the same where also the ionized jet is centered on. Considering our spatial resolution of $\sim$97\,au, this innermost strongest peak should host at least a binary, if not an even higher order multiple system. A potential candidate for that is the proposed additional protostar $\sim 0.18''$ south of the main HW2 source (maser group R4, \citealt{torrelles2011}), roughly within our central mm continuum peak.

Regarding other suggested protostellar objects (e.g., \citealt{curiel2002,martin-pintado2005,comito2007}), we clearly see significant additional 1.3\,mm continuum emission peaks in the central elongated disk-like structure (e.g., Fig.~\ref{cont} right panel) that we will discuss in the following section.

\subsection{Fragmentation of a close to edge-on high-mass disk}
\label{disk-discussion}

The flattened and elongated structure of the extended central 1.3\,mm emission is suggestive of an almost edge-on disk configuration. In such a configuration, one can ask whether the additional emission peaks in the structure are indeed fragments that may host additional protostars or whether they may even represent rings like those found towards low-mass disks (e.g., \citealt{andrews2018}) and also suggested for one high-mass source (G17, \citealt{maud2019}). As outlined in section \ref{disk} and Fig.~\ref{tau}, most of the extended disk-like structure is optically thin, hence really tracing all compact disk structure along the line of sight.

Most low-mass disk studies at (sub)mm wavelengths typically targeted disks that are not edge-on but have inclination angles that allow to resolve the structures (e.g., \citealt{andrews2018,manara2023}). However, notable exceptions exist. While \citet{villenave2020} studied 12 low-mass edge-on disks, another particularly interesting target is the low- to intermediate-mass edge-on disk IRAS\,23077+6707 \citep{lovell2025,monsch2026}. Although most of the edge-on disks in \citet{villenave2020} have relatively flat intensity profiles without much sub-structures, IRAS\,23077+6707 has several emission peaks along the disk \citep{lovell2025}, similar to the high-mass case CepA studied here. In addition to this, with an extent of $\sim 6''$ at mm wavelengths, and the given distance uncertainties (150--300\,pc, \citealt{lovell2025,monsch2026}), its size between 900 and 1800\,au is comparable to the CepA elongated disk-like structure size. Regarding the observed asymmetric intensity peaks, \citet{lovell2025} discuss several potential origins like eccentricity, a misaligned inner disk, an arc or a spiral. However, they cannot clearly distinguish any of the options based on their given data. Since high-mass star-forming regions are known to form multiple objects with often high degrees of multiplicity (e.g., \citealt{offner2023}), the disk asymmetries may potentially also stem from a fragmented disk where the disk fragments may host seeds for additional protostars within the system. 

An interesting comparison work to these CepA data, are the analysis of 3D radiation-hydrodynamic simulations that \citet{ahmadi2019} post-processed  with radiation transport and interferometer simulators to create realistic 1.3\,mm continuum and CH$_3$CN$(12_k-11_k)$ spectral line data for different ALMA and NOEMA configurations. Starting with a 200\,M$_{\odot}$ mass reservoir and infalling envelope at an outer radius of 0.1\,pc, these simulations form a high-mass disk that fragments and forms almost spiral like structures. A snapshot at 12\,kyr is taken when the central protostar has grown to 10\,M$_{\odot}$.

Figure \ref{ahmadi} (left panels) presents the same simulations and corresponding modeled 1.37\,mm continuum data shifted to a distance of 800\,pc (\citealt{ahmadi2019}, corresponding to their Fig.~3). These properties align with those of CepA and our NOEMA observations at a distance of $\sim$700\,pc. Also the linear extent of the simulated disk larger than 1000\,au compares well to our data. We present the 10 and 80\,deg simulations that on the one hand give a good overview of the general disk and fragment structure (10\,deg inclination in Fig.~\ref{ahmadi}), and on the other hand outline how such structures look in an almost edge-on configuration (80\,deg inclination in Fig.~\ref{ahmadi}). The 10\,deg representation clearly shows the multiple fragments that formed within the high-mass disk. Interestingly, when rotating that to the more edge-on 80\,deg inclination configuration, one sees an elongated structure with several emission peaks. Some of these emission peaks are close to the mid-plane of the elongated emission, but one also sees some fragments that are still offsets a bit to the top and bottom of the configuration. In addition to that, the pv-diagrams presented in \citet{ahmadi2019}, their Fig.~7, also show emission in the other two non-disk quadrants, similar to our CepA pv-diagrams in Fig.~\ref{pv}. These additional structures are from the infalling envelope in \citet{ahmadi2019}. More detailed analysis of infall kinematics in CepA will be conducted in future studies. 

For comparison, we investigate how ring-like structures would appear in an edge-on configuration. For that purpose, we took the fragmented disk from \citet{ahmadi2019} and azimuthally smoothed the emission. In that way, at the radii where originally the fragments are located, one now finds rings. The right panels in Figure \ref{ahmadi} present the corresponding 1.37\,mm emission again at 10 and 80\,deg inclination (top and bottom panels, respectively). One finds that, in contrast to the fragmented disk, in the ring-like configuration at 80\,deg inclination the emission is much smoother without any pronounced additional emission peaks along the disk extent. We note that different configurations of ring-like structures with different sizes and relative contrast may result in less smooth edge-on emission structures. However, creating an exact model that could reproduce our observations is beyond the scope of this paper.

Comparing these simulations to our observations, qualitatively speaking, the almost edge-on 80\,deg inclination angle simulations of the fragmented disk agree surprisingly well with our data, whereas the ring-configuration in its smooth edge-on emission does not resemble our observations. Regarding the fragmented disk, we not only find similar sizes for the entire structure and similar numbers of emission peaks along the fragmented disk, but also the fact that some modeled fragments are closer to the projected mid-plane whereas other are at slightly offset positions corresponds in our data to core \#5 and the sub-structure south labeled in Fig.~\ref{cont}. These simulations were conducted with much more general application in mind \citep{ahmadi2019}, yet it is remarkable that they apply to the case of CepA without modifications. 

Therefore, although we cannot exclude other possibilities for the observed sub-structures, like ring-structures with different ring sizes or contrasts as mentioned above, the qualitative comparison of our 1.3\,mm continuum data with the simulations from \citet{ahmadi2019} support the idea that we may be observing a fragmenting high-mass disk in an almost edge-on configuration. This scenario is further supported by the Toomre $Q$ analysis that finds low $Q$ values towards the sub-structures which are indicative of instabilities and hence fragmentation. If one now compares the flux densities (Fig.~\ref{cont}), optical depths (Fig.~\ref{tau}) and temperatures (Fig.~\ref{trot}), one finds that all are in similar ranges for the four disk fragments east1, east2, west1 and west2. Hence, in the fragmented disk scenario, the disk fragments should have also similar masses, consistent with theoretical expectations (e.g., \citealt{oliva2020,mignon-risse2023}).

Another interesting aspect is that some of the fragments, in particular west1 and west2, appear elongated roughly perpendicular to the disk (Fig.~\ref{cont}). Since the synthesized beam has a position angle of $95^{\circ}$, this is not an observational artifact but should be real. The direction of this elongation corresponds also roughly to the direction of the magnetic field \citep{vlemmings2010}. These elongations require further analysis and modeling in future work.
\section{Conclusions}

The high-angular resolution observations ($\leq 0.2''$) of one of the closest high-mass star-forming regions Cepheus A HW2 resolve the inner disk and outflows at an unprecedented spatial resolution. While we find several protostellar condensations within the field of view, the central disk-like structure shows several emission peaks roughly along its major axis and also slightly offset from it. Except for the central peak, the  1.3\,mm continuum emission is largely optically thin and hence traces all compact disk emission along the line of sight. The strongly flattened structure is suggestive of an almost edge-on disk configuration.

Comparing our data to radiative-transfer and post-processed 3D radiation hydrodynamic simulations we find that our data are most consistent with a fragmented high-mass accretion disk, but are not well represented by ring-structures within a disk. This fragmentation inference is also supported by a Toomre $Q$ analysis of the structure with low $Q$ values towards the 1.3\,mm continuum emission in the extended disk structure.

The CO and SiO spectral line data confirm the previously suggested second outflow emanating from the central peak position in east-west direction (different to the well-known northeast-southwest jet). With a linear resolution of $\sim$97\,au, this unresolved central peak most likely hosts a binary or even higher order multiple system.

A kinematic analysis of the dense gas tracer CH$_3$CN is complicated by the fact that at least the innermost emission close to the central peak and outflow-driving source is clearly affected by the outflows themselves. Hence, we cannot attribute the CH$_3$CN emission unambiguously to the disk structure. A pv-analysis reveals features consistent with disk emission, but also strong emission in the other pv-quadrants that may be caused potentially by infall motions. A more detailed kinematic analysis will be kept for future studies where it is planned to combine these long-baseline data with the original data from the CORE project that cover shorter baselines as well as short spacing information \citep{beuther2018,gieser2021,ahmadi2023}.

\begin{acknowledgements}
The authors are grateful to the staff at the NOEMA
observatory for their support of these observations.  This work is based on observations carried out under project number W22AL001. IRAM is supported by INSU/CNRS (France), MPG (Germany) and IGN (Spain). RK acknowledges financial support via the Heisenberg Research Grant funded by the Deutsche Forschungsgemeinschaft (DFG, German Research Foundation) under grant no.~KU 2849/9, project no.~445783058. DS was funded by the Deutsche Forschungsgemeinschaft (DFG, German Research Foundation) – project number: 550639632. A.S.-M. acknowledges support by the grant PID2023-146675NB-I00 (MCI-AEI-FEDER, UE). This work is also partially supported by the Spanish program Unidad de Excelencia Mar\'ia de Maeztu CEX2020-001058-M, financed by MCIN/AEI/10.13039/501100011033, and by the MaX-CSIC Excellence Award MaX4-SOMMA-ICE. RGM acknowledges support from UNAM-DGAPA-PAPIIT projects IN105225. REP is supported by a Discovery grant from the National Science and Engineering Research Council (NSERC) of Canada. AP acknowledges financial support from the UNAM-PAPIIT IN120226 grant, and the Sistema Nacional de Investigadores of SECIHTI, M\'exico.
\end{acknowledgements}  


\begin{appendix}
\section{Additional figures}

\begin{figure}[ht]
	\includegraphics[width=0.99\linewidth ,keepaspectratio]{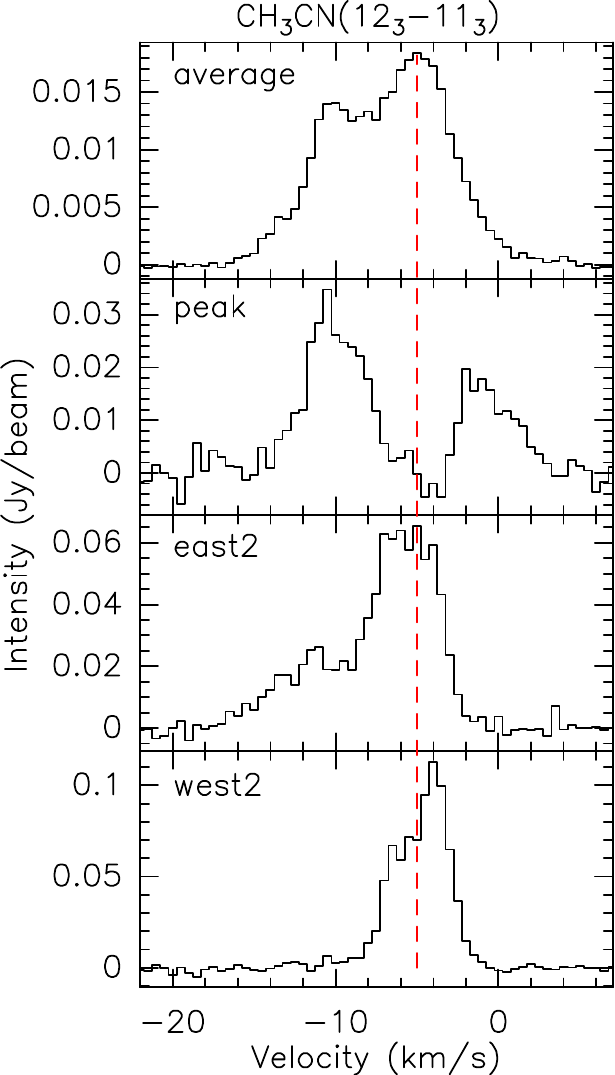}
	\caption{CH$_3$CN$(12_3-11_3)$ spectra at different positions in CepA. The top-spectrum shows the integrated emission of the central structure. The following three panels show the spectra extracted towards the central peak and the positions east2 and west2 (Fig.~\ref{intensitycut}), respectively. The approximate $v_{\rm lsr}$ of $\sim -5$\,km\,s$^{-1}$ is marked by the dashed red line.}
	\label{spectra_ch3cn}
\end{figure}
 
\begin{figure*}[ht]
	\includegraphics[width=0.99\linewidth ,keepaspectratio]{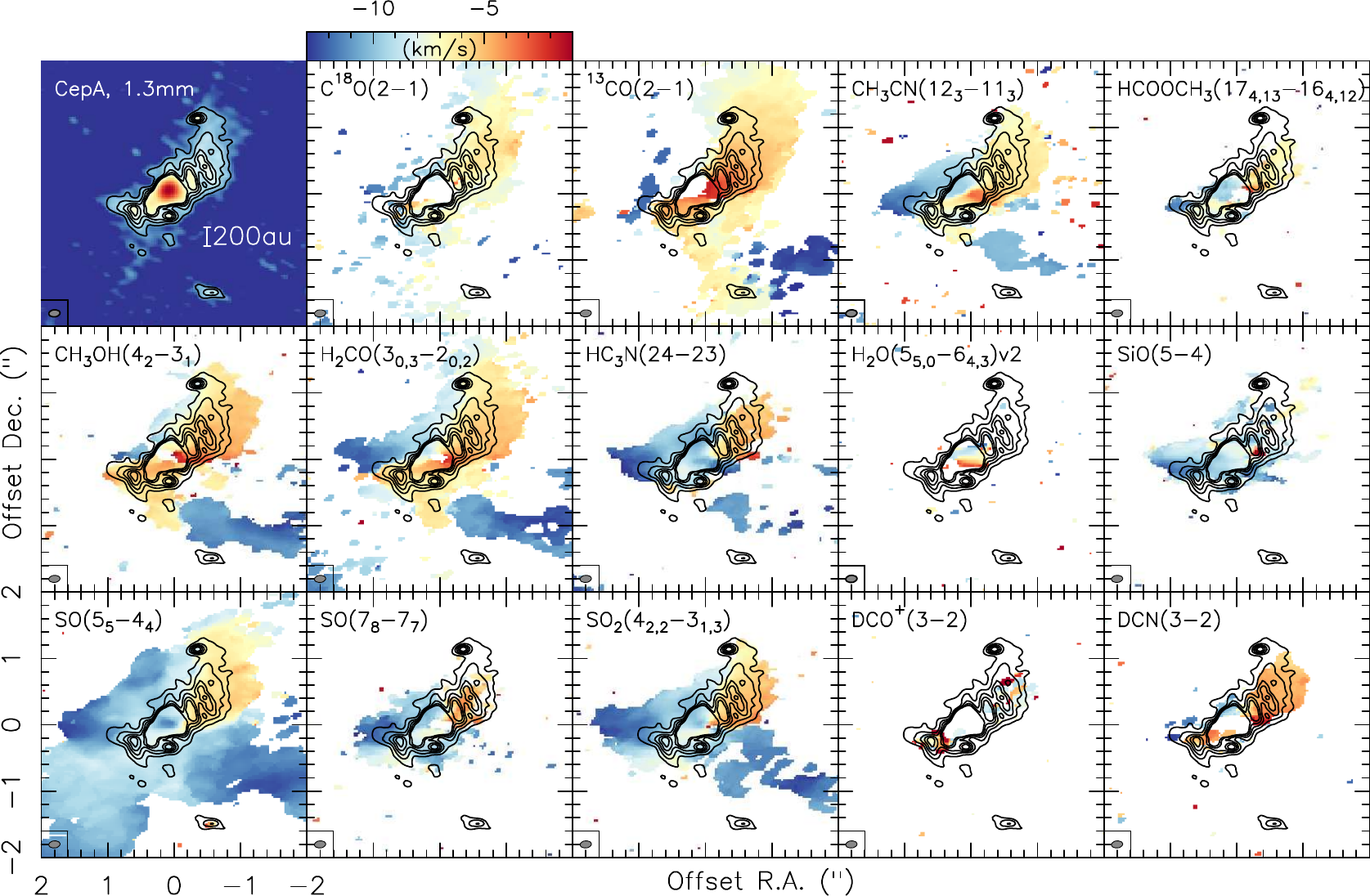}
	\caption{First moment maps (intensity-weighted peak velocities) for selected molecules (Fig.~\ref{mom0} and \ref{rms}). The top-left panel shows in color and contours the 1.3\,mm continuum emission, whereas in all other panels the color presents the moment 1 maps created above the $5\sigma$ values. Integration regimes for most panels are from $-$13 to $-$1\,km\,s$^{-1}$ (color scale above C$^{18}$O panel), only for CH$_3$CN, HCOOCH$_3$ and H$_2$O slightly different velocity regimes were chose ([$-15$,3], [$-13$,3], [$-30$,7], respectively, for H$_2$O see also Fig.~\ref{overlay_h2o}). The contours again show the 1.3\,mm continuum, contour levels are always in $4\sigma$ steps. The synthesized beam is shown in all panels, the the top-left panel also presents a scale-bar.}
	\label{mom1}
\end{figure*}

\begin{figure*}[ht]
	\includegraphics[width=0.99\linewidth ,keepaspectratio]{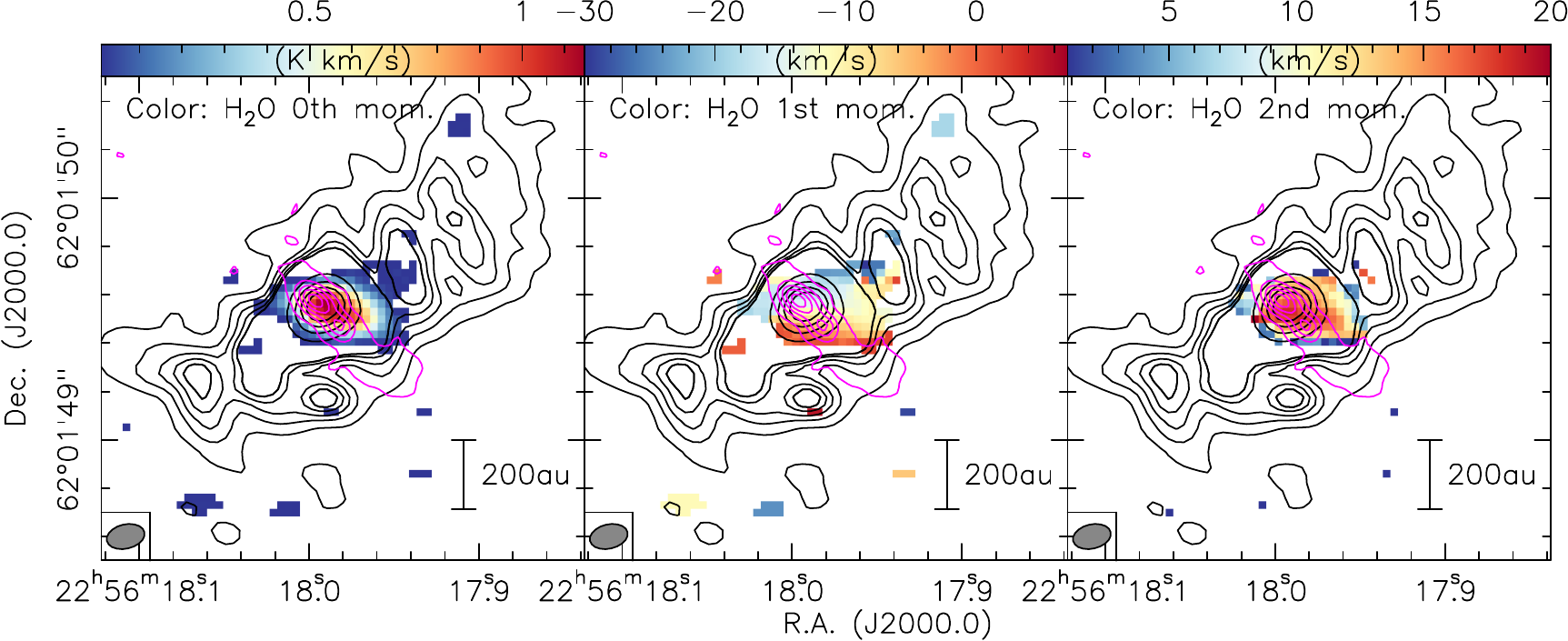}
	\caption{Overlay of H$_2$O and 1.3\,mm data. The three panels show in color the 0th, 1st and 2nd moments of H$_2$O emission, respectively. The contours outline the 1.3\,mm data from 4$\sigma$ to $20\sigma$ in $4\sigma$ steps and then continue in 100$\sigma$ steps. The magenta contours show the 1.3\,cm continuum jet \citep{torrelles1996}. Synthesized beams and scale-bars are presented as well.}
	\label{overlay_h2o}
\end{figure*}

\end{appendix}

%

\end{document}